\documentclass[sigconf]{acmart}
\usepackage{array}
\usepackage{graphicx}
\usepackage{clrscode}
\usepackage{balance}
\usepackage{subfigure}
\usepackage{multirow}
\usepackage{multicol}
\usepackage{float}
\usepackage{color}
\usepackage{xcolor}
\usepackage{amsopn}
\usepackage{mathrsfs}
\usepackage{mathtools}
\usepackage{amsmath}
\usepackage{booktabs}
\usepackage{arydshln}
\usepackage{hyperref}
\usepackage{blkarray}
\usepackage{enumerate}
\usepackage{courier}
\usepackage{mathrsfs}
\usepackage{rotating}
\usepackage{bm}
\usepackage{subfigure}
\usepackage{array}
\usepackage[lined,boxruled,commentsnumbered,linesnumbered]{algorithm2e}
\begin{document}
\title{Visually Explainable Recommendation}

\author{Xu Chen}
\affiliation{
 \institution{School of Software}
  \institution{Tsinghua University}
}
\email{xu-ch14@mails.tsinghua.edu.cn}
\author{Yongfeng Zhang}
\affiliation{
 \institution{Department of Computer Science}
  \institution{Rutgers University}
}
\email{yongfeng.zhang@rutgers.edu}
\author{Hongteng Xu}
\affiliation{
 \institution{Department of ECE}
  \institution{Duke University}
}
\email{hongteng.xu@duke.edu}
\author{Yixin Cao}
\affiliation{
 \institution{Department of Computer Science}
  \institution{Tsinghua University}
}
\email{caoyixin2009@163.com}
\author{Zheng Qin}
\affiliation{
 \institution{School of Software}
  \institution{Tsinghua University}
}
\email{qingzh@mail.tsinghua.edu.cn}
\author{Hongyuan Zha}
\affiliation{
 \institution{College of Computing}
  \institution{Georgia Institute of Technology}
}
\email{zha@cc.gatech.edu}

\begin{abstract}
Images account for a significant part of user decisions in many application scenarios, such as product images in e-commerce, or user image posts in social networks. It is intuitive that user preferences on the visual patterns of image (e.g., hue, texture, color, etc) can be highly personalized, and this provides us with highly discriminative features to make personalized recommendations.

Previous work that takes advantage of images for recommendation usually transforms the images into latent representation vectors, which are adopted by a recommendation component to assist personalized user/item profiling and recommendation. However, such vectors are hardly useful in terms of providing visual explanations to users about why a particular item is recommended, and thus weakens the explainability of recommendation systems. 

As a step towards explainable recommendation models, we propose \textit{visually explainable recommendation} based on attentive neural networks to model the user attention on images, under the supervision of both implicit feedback and textual reviews. By this, we can not only provide recommendation results to the users, but also tell the users why an item is recommended by providing intuitive visual highlights in a personalized manner. Experimental results show that our models are not only able to improve the recommendation performance, but also can provide persuasive visual explanations for the users to take the recommendations.

\end{abstract}


\maketitle

\section{Introduction}
Recommender systems have been major building blocks for many online applications, which provide users with personalized suggestions. 
To capture users' personalized preferences as comprehensive as possible, recommender systems have integrated a wide range of information sources in the modeling process. 

Image -- which widely exists in e-commerce, social networks, and many other applications -- is one of the most important resources integrated into modern recommender systems.
Previous work that leverage images for personalized recommendation usually transform images into embedding vectors, which are then incorporated with collaborative filtering (CF) for improving the performance. For example, McAuley et al \cite{mcauley2015image} adopted neural networks to transform images into feature vectors, and used the vectors for product style analysis and recommendation; He et al \cite{he2016vbpr} further extended the approach to pair-wise learning to rank for recommendation; Geng et al \cite{geng2015learning} adopted image features for recommendation in a social network setting; and Wang et al \cite{wang2017your} extracted image features with neural network for point-of-interest recommendation.

Though the recommendation performance has been improved by incorporating image representation vectors extracted with (convolutional) neural networks, the related work has largely ignored an important advantage of leveraging images for recommendation -- its ability to provide intuitive visual explanations. 
This is because by transforming the whole image into a fixed latent vector, the images become hardly understandable for users, which makes it difficult for the model to generate visual explanations to accompany certain recommendations.

\begin{figure}[t!]
\centering
\setlength{\fboxrule}{0.pt}
\setlength{\fboxsep}{0.pt}
\fbox{
\includegraphics[width=0.95\linewidth]{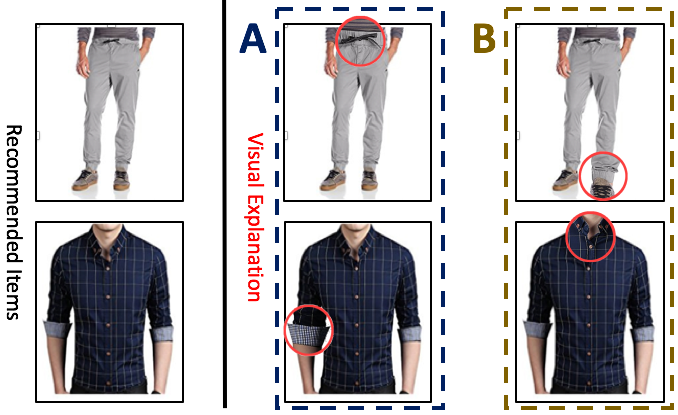}
}
\caption{Personalized visual explanations for the recommended items.
Here, different parts of the images are magnified in circle to provide intuitive explanations for the corresponding user.
Specifically, the pants and T-shirt are both recommended to user A and B. For user A, the waistband of the pants and the cuff of the T-shirt are highlighted to tell the user why these items are recommended, while for user B, the pants turnup and the T-shirt collar are magnified as personalized explanations.}
\label{intro}
\vspace{-10pt}
\end{figure}

Researchers have pointed out long ago that providing appropriate explanations is beneficial to recommender systems in terms of recommendation persuasiveness, satisfaction, effectiveness, and scrutability, etc \cite{herlocker2000explaining,tintarev2007survey}. Existing explainable recommendation models usually interpret the recommendations based on user reviews~\cite{mcauley2013hidden,zhang2014explicit,seo2017interpretable}. 
However, \textit{``a picture may paint a thousand words''}, textual features may be less intuitive compared with the visual ones. As exampled in Figure \ref{intro}, the magnified region of the pants image can intuitively tell user A \textit{``the color, position, style, ... of the waistband''}, while describing them in text may cost a lot of words, and fail to provide intuitive understandings.

Motivated by the desire to fill the gap, in this paper, we propose to provide \textit{personalized visual explanations} with image highlights (see Figure \ref{intro}) in the context of image-based recommendation. 
The key building block of our model is the integration of attention mechanism and collaborative filtering.
Specifically, we first design a basic method to model user attention on the product images, and use the learned attention weights to provide visual explanations. 
Then, to capture more comprehensive preferences, we further extend the basic model by introducing user reviews as an additional weak supervision signal. 
Compared with existing methods, our approaches not only improve the recommendation performance, but also generate visual explanations for the recommended items.

\textbf{Contributions.} In summary, the main contributions of this work include:
\begin{itemize}
\item We propose \textit{visually explainable recommendation} to explain recommendations from the visual perspective, which, to the best of our knowledge, is the first time in the research of personalized recommendation.

\item We design two types of neural attentive models to discover user visual preference, with the supervision of implicit feedback as well as textual reviews. With the learned attention weights, we can readily generate visually explainable recommendations. 

\item We conduct experiments to verify the effectiveness of our proposed models for Top-N recommendation as well as review prediction. Further more, we release a collectively labeled dataset to quantitatively evaluate our generated visual explanations, and also we present example analysis to highlight the intuitions of the visual explanations in a qualitative manner.
\end{itemize}

In the following part of the paper, we first introduce the related work in section 2, and then provide the problem formalization in section 3.
Section 4 illustrates the details of our proposed model, and in section 5, we verify the effectiveness of our approach with experimental results. Conclusions and outlooks of this work are presented in section 6.

\section{Related Work}
There exist two main research lines related to our work -- explainable recommendation and the integration of visual features into recommendation systems. We present brief reviews for these two research lines in the following.
\subsection{Explainable Recommendation}
Many models have been proposed in the recent years to provide explainable recommendations~\cite{zhang2017explainable}. 
In specific, McAuley et al~\cite{mcauley2013hidden} aligned user (or item) latent factors in matrix factorization (MF) with topical distribution in latent dirichlet allocation (LDA)~\cite{blei2003latent} for joint parameter optimization under the supervision of both score ratings and textual reviews, and thus the user preferences are explained by the learned topical distributions.
To explain finer-grained user preference, Zhang et al~\cite{zhang2014explicit} translated user reviews into feature-opinion pairs, and then leveraged multi-matrix factorization to discover user preferences as well as item qualities on the feature-level. 
Wu et al~\cite{wu2016explaining} designed an additive co-clustering model based on Gaussian and Poisson distributions to explain recommendations by jointly optimizing user reviews and ratings, while Heckel et al~\cite{heckel2016interpretable} generated interpretable recommendations by identifying overlapping co-clusters of clients and products based on implicit feedback.
By modeling aspects in user reviews, He et al~\cite{he2015trirank} devised a graph algorithm called \textit{TriRank} for providing recommendations with better explainability and transparency, and to leverage user opinions as well as social information, Ren et al~\cite{ren2017social} introduced the concept of \textit{viewpoints} (a tuple of concept, topic, and sentiment label), and proposed a probabilistic graphical model based on viewpoints to provide explainable recommendations.
Seo et al~\cite{seo2017interpretable} utilized attention mechanism to find both local and global preference information in the textual reviews to explain the users' rating behaviors. 

Different from the above methods that are mainly based on user reviews, we explore to capture users' visual preferences, and provide explainable recommendations from a new visual perspective.

\subsection{Image-based Recommendation}
Recently, there is a trend to incorporate visual features into the research of personalized recommendation. 
Specifically, McAuley et al~\cite{mcauley2015image} introduced the concept of visual recommendation into e-commerce, and released a large dataset for this task.
He et al~\cite{he2016vbpr} represented each product image as a fixed length vector, and infused it into the bayesian personalized ranking (BPR) framwork~\cite{rendle2009bpr} to improve the performance of Top-N recommendation. 
To make use of both visual- and textual- features, Cui et al \cite{cui2016visual} integrated the product images and item descriptions together to make dynamic Top-N recommendation.
Liu et al~\cite{liu2017deepstyle} adopted neural modeling based on product images to model the style of items, which led to improved recommendation performance.
Wang et al~\cite{wang2017your} introduced image features into point-of-interest (POI) recommendation, and proposed a graphical framework to model visual content in the context of POI recommendation.
Shankar~\cite{shankar2017deep} designed a unified end-to-end neural model (named VisNet) to build a large scale visual recommendation system for e-commerce.
Chen~\cite{chen2017attentive} introduced the attention mechanism into CF to model both item- and component-level implicit feedback for multimedia recommendation.

Generally, these methods mainly focus on how to leverage image features to enhance the recommendation performance, however, we want to discover users' personalized visual preferences, and more importantly, to provide visually explainable recommendations.

%We slso notice that the model designed in our work is somewhat related to the architecture proposed for ``image captioning''~\cite{vinyals2015show,xu2015show} in the computer vision (CV) community, however, we focus on a different problem, and our model is actually a ``multi-task learning'' framework, which is different from their ``single-task'' model.

\section{Problem Definition}
Suppose there are $N$ users $\bm{u} = \{u_1,u_2,...,u_N\}$ and $M$ items $\bm{v} = \{v_1,v_2,...,v_M\}$ in our system. 
Each item $j \in \bm{v}$ has an image $G_j$, and we extract item $j$'s visual features as $\bm{f}_j = \{\bm{f}_j^1, \bm{f}_j^2, ..., \bm{f}_j^h\}$ from $G_j$ leveraging existing methods such as deep convolutional neural networks~\cite{simonyan2014very}, where $\bm{f}_j^k \in R^D~(k\in \{1,2,...,h\})$ is a $D$ dimensional vector corresponding to a spatial region of image $G_j$, and $h$ is the number of different regions.
Let $\bm{w}_{ij} = \{{w_{ij}^1}, {w_{ij}^2}, ..., {w_{ij}^{l_{ij}}}\}$ ($i\in \bm{u}, j \in \bm{v}$) be the textual review of user $i$ on item $j$, where $w_{ij}^t~(t\in \{1,2,...,l_{ij}\})$ is the $t$-th word, and $l_{ij}$ is length of the review.

Given visual features $\bm{F} = \{\bm{f}_j | j \in \bm{v}\}$, user reviews $\bm{W} = \{\bm{w}_{ij} | i\in \bm{u}, j \in \bm{v}\}$ and user-item historical interaction records, our task is to learn a recommendation model to (1)~generate top-$n$ personalized items as recommendations for a target user, and (2)~provide explanations for these recommended items based on $\bm{F}$. For easy reference, we list the notations used throughout the paper in Table~\ref{tab-notation}.

\section{\mbox{Visually Explainable Recommendation}}
In this section, we first propose a base model to attentively incorporate image features into collaborative filtering (CF) to provide visually explainable recommendations.
And then, an advanced model is designed to leverage user textual reviews to further enhance the recommendation performance as well as interpretability.

\begin{figure*}[t!]
\centering
\setlength{\fboxrule}{0.pt}
\setlength{\fboxsep}{0.0pt}
\fbox{
%\subfigure[Collaborative filtering as neural network]{
%\includegraphics[width=0.25\linewidth,height=0.2\linewidth]{fig/cf.png}\label{feature-level}\label{intro-rnn}
%}
\subfigure[Visually explainable collaborative filtering (VECF)]{
\includegraphics[width=0.43\linewidth,height=0.28\linewidth]{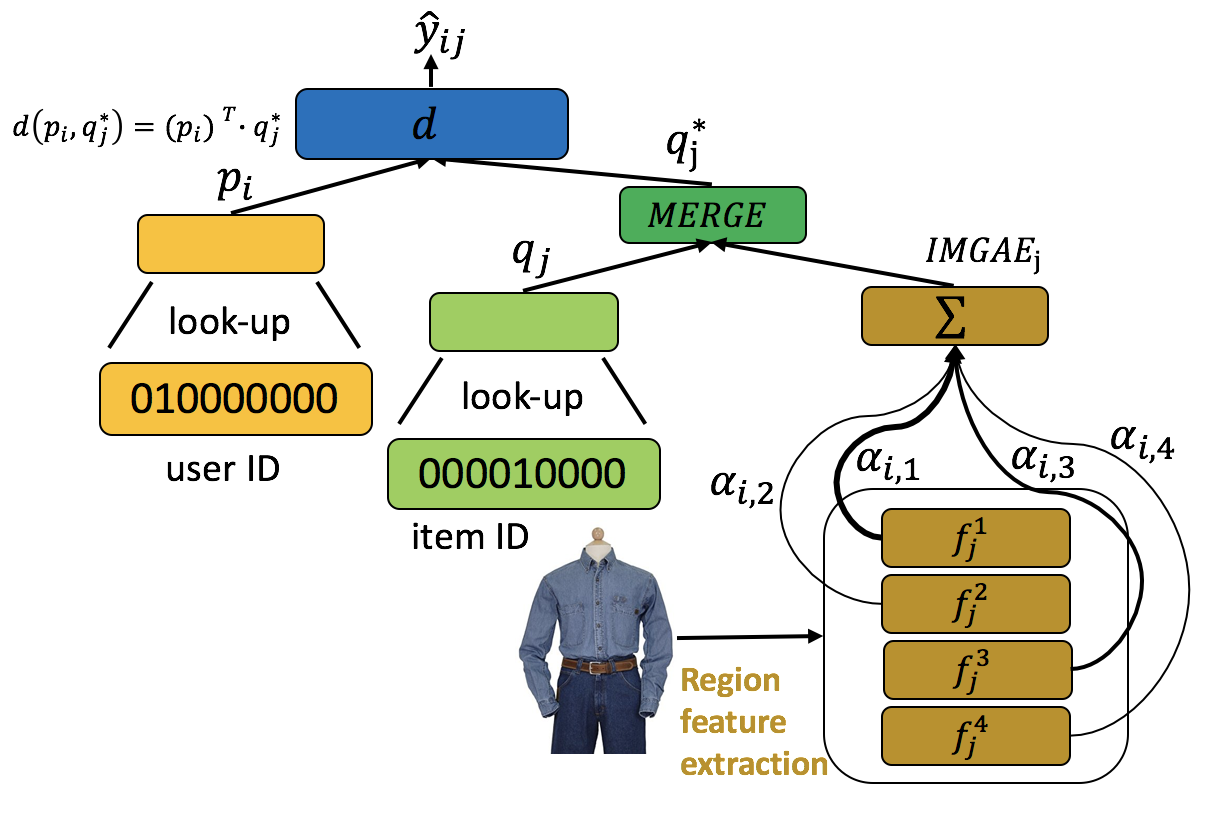}\label{vec-nn}
}
\hspace{.15in}
\subfigure[Review-enhanced visually explainable collaborative filtering (Re-VECF)]{
\includegraphics[width=0.43\linewidth,height=0.28\linewidth]{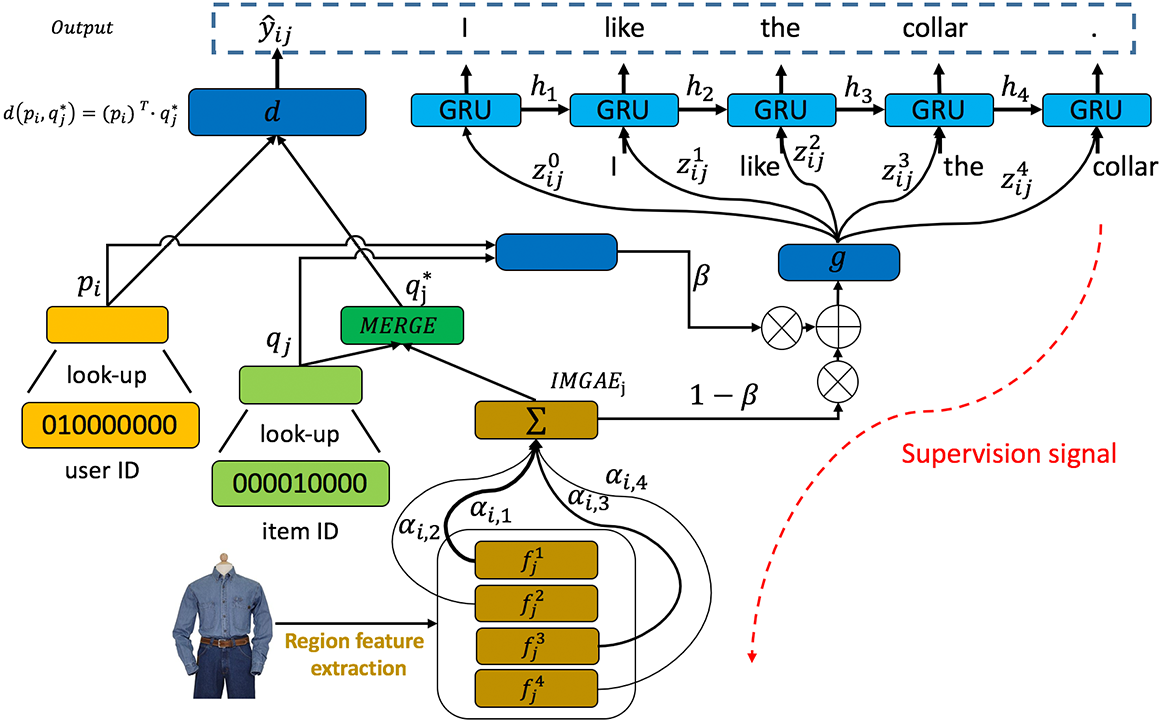}\label{re-vec-nn}
}
}
\vspace{-5pt}
\caption{
(a) Our base model. The regional features are attentively merged into an image representation, which is then multiplied with the user embedding to predict the final result. The learned attention weights are used to generate visual explanations.
(b) The review enhanced model. User reviews are incorporated into the architecture to provide more informative signals to enhance the recommendation performance as well as the interpretability.
}
\vspace{-10pt}
\end{figure*}

\subsection{The Base Model}
Intuitively, when browsing a product image, users often pay more attention to the regions related to their interests, while the attention cased on other parts may be relatively less.
For example, a user who wants to buy a round neck T-shirt may care more about the collar relevant regions compared with other areas. 
To model such human senses, we design \textbf{V}isually \textbf{E}xplainable \textbf{C}ollaborative \textbf{F}iltering (\textbf{VECF} for short) based on attention mechanism to discover user's region-level preferences, and use the learned attention weights to explain recommendations from the visual perspective.

In the following, we first briefly describe the method for image feature extraction, and then illustrate the model details.

\subsubsection{\textbf{Image feature extraction}}
Considering the efficiency for practical applications, in our model, we pre-extract the regional visual features of images, which is similar to many previous work~\cite{chen2017attentive,he2016vbpr,liu2017deepstyle}. 
Specifically, we feed each product image into the pre-trained VGG-19~\cite{simonyan2014very} architecture, and use the output of $conv5$ layer as the final representation. This $14\times 14\times 512$ feature map can be seen as 196 feature vectors ($h=196$) of 512-dimension ($D=512$), corresponding to 196 different square regions of the image.

This type of pre-processing is essentially equivalent to training an end-to-end model by fixing the pre-trained parameters as in VGG. 
If more computational resources are available, we can also free the VGG component to achieve a totally end-to-end model for visual feature learning.
%and investigate whether can we obtain more effective visual region features. 
%We leave this exploration as one of our future work.

\begin{table}[!t]
  \centering
  \caption{Notations and descriptions.}
  \vspace{-10pt}
  \label{tab-notation}%
  \begin{tabular}{cp{0.56\columnwidth}}
    \hline\hline
    Notations & Descriptions \\
    \hline
    $\bm{u}$  & The set of $N$ users $\{u_1,u_2,...,u_N\}$.\\
    $\bm{v}$  & The set of $M$ items $\{v_1,v_2,...,v_M\}$.\\\hline
    $\bm{p}_i, ~\bm{q}_j, ~K$ & The user and item embeddings, and their corresponding dimension.\\
    $\bm{f}_j, ~D, ~\bm{IMAGE}_j, ~\bm{F}$  & Regional feature set $\{\bm{f}_j^1, \bm{f}_j^2, ..., \bm{f}_j^h\}$ for item $j$, the dimension of each region feature, the merged region feature for item $j$, and the set of all visual features $\{\bm{f}_j | j \in \bm{v}\}$\\
    $\bm{w}_{ij}, ~\bm{c}_{ij}^t, ~\bm{W}$ & The word list in the review of user $i$ on item $j$: $\{{w_{ij}^1}, {w_{ij}^2}, ..., {w_{ij}^{l_{ij}}}\}$, the one-hot format of word ${w_{ij}^t}$, and the set of all reviews $\{\bm{w}_{ij} | i\in \bm{u}, j \in \bm{v}\}$.\\\hline
    $\alpha_{i,j,1},~\alpha_{i,j,2},...,~\alpha_{i,j,h}$ & The visual attention weights of user $i$ on item $j$'s product image.\\
    $\bm{w}^{u}_{att}, ~\bm{w}^{r}_{att}, ~b_{att}$ & The weighting and bias parameters for visual attention weights.\\
    $\bm{h}$&The hidden states in GRU.\\
    $\bm{W}^z_{g}, ~\bm{W}^r_{g}, ~\bm{W}^h_{g}$ &The parameter matrices for the input word embedding of GRU.\\
    $\bm{U}^z_{g}, ~\bm{U}^r_{g}, ~\bm{U}^h_{g}$ &The parameter matrices for the hidden state of GRU.\\
    $\bm{V}^z_{g}, ~\bm{V}^r_{g}, ~\bm{V}^h_{g}$ &The parameter matrices for the visual features in our revised GRU.\\
    $\bm{b}^z_g, ~\bm{b}^r_g, ~\bm{b}^h_g \in R^{Z}$ &The bias vectors of GRU.\\
    $\bm{z}_{ij}^t$& Context vector for generating the $t$-th word in the review of user $i$ on item $j$.\\
    $\bm{W}^{u}_{c},\bm{W}^{i}_{c},\bm{W}^{img}_{c},\bm{w}^{h}_{c},\bm{b}_{c}$& The weighting and bias parameters used to derive context vectors.\\\hline
    $\bm{E}$& The word embedding matrix.\\
    $N^w$&The vocabulary size.\\
    $O$&The word embedding dimension.\\
    $Z$&The hidden state dimension.\\
    $\sigma, \text{tanh}, g$&The sigmoid, hyperbolic tangent and ReLU activation functions.\\
    \hline\hline
    \end{tabular}%
\vspace{-10pt}
\end{table}%

\subsubsection{\textbf{Visually Explainable Collaborative Filtering (VECF)}}
The overall design principle of our base model is shown in Figure~\ref{vec-nn}. Suppose $\bm{p}_{i}\in R^K$, $\bm{q}_{j}\in R^K$ are the embeddings of user $i$ and item $j$, respectively. The visual feature of $j$ is $\bm{f}_j = \{\bm{f}_j^1, \bm{f}_j^2, ..., \bm{f}_j^h\}$, where $\bm{f}_j^k \in R^D~(k\in \{1,2,...,h\})$ is a $D$-dimension vector corresponding to a spatial region of $j$'s image, and $h$ is the number of different regions. 
We compute image $j$'s global image feature with:
\begin{equation}
\bm{IMAGE}_{j} = \sum\nolimits_{k=1}^h \alpha_{i,j,k} \cdot \bm{f}_j^k
\end{equation}
where $\alpha_{i, j, k}$ is the attention weight jointly determined by the current user $i$ and the regional feature $\bm{f}_j^k$, which is: 

\begin{equation}
\begin{aligned}
&a_{i, j, k} = g\big((\bm{w}^{u}_{att})^T\cdot \bm{p}_{i} + (\bm{w}^{r}_{att})^T\cdot \bm{f}_j^k + b_{att}\big)\\\label{image_att}
&\alpha_{i,j,k} = \frac{a_{i,j,k}}{\sum_{\kappa=1}^h a_{i, j, \kappa}}
\end{aligned}
\end{equation}
where $\bm{w}^{u}_{att} \in R^K$, $\bm{w}^{r}_{att}\in R^D$ and $b_{att} \in R$ are the parameters to learn, $g(x) = \max(0, x)$ is the Rectified Linear Unit (ReLU)~\cite{maas2013rectifier} active function.

Given item $j$'s global image feature $\bm{IMAGE}_{j}$, the final item embedding $\bm{q}_j^*$ can be computed as:
\begin{equation}
\bm{q}_j^* = MERGE(\bm{q}_j,~\bm{IMAGE}_j)
\end{equation}
where $MERGE(\cdot)$ is a function that merges the image representation and the item's latent embedding. Here, we implement it as a simple element-wise multiplication, and other choices including element-wise addition and vector concatenation have also been tested, but they lead to unfavored performance.

When making predictions, we feed the user embedding $\bm{p}_i$ and the final item embedding $\bm{q}_j^*$ into a function:
\begin{equation}
\hat y_{ij} = PREDICT(\bm{p}_i,~\bm{q}_j^*)
\end{equation}
where $PREDICT(\cdot)$ is an arbitrary prediction function, or even a prediction neural network as in~\cite{he2017neural}.
Here, we choose the sigmoid inner product $\hat y_{ij} = \sigma(\bm{p}_i^T\cdot \bm{q}_j^*)$ as a specific implementation, because it gives us better training efficiency on our large-scale data. However, it is not necessarily restricted to this function and many others can be used in practice according to the application domain.

At last, we adapt binary cross-entropy as the loss function for model optimization, and the objective function to be maximized is:
\begin{equation}
\begin{aligned}
l_{1}&= \log~\prod_{(i,j)}({\hat y_{ij}})^{y_{ij}}{(1-\hat y_{ij})}^{1-y_{ij}} - \lambda ||\Theta||^2_F\\
&=\sum_{i\in \bm{u}}\sum_{j \in \bm{v}_{+}^i}\log \hat y_{ij} + \sum_{i\in \bm{u}}\sum_{j \in \bm{v}/\bm{v}_{+}^i} \log(1-\hat y_{ij})- \lambda ||\Theta||^2_F
\end{aligned}
\label{formular-loss}
\end{equation}
where $\Theta$ is the model parameter set, $y_{ij}$ is the ground truth that would be 1 if user $i$ purchased item $j$, and 0 otherwise.
$\bm{v}$ is the set of all items, and $\bm{v}_{+}^i$ is the set of items that $i$ has purchased before. 
Corresponding to each positive instance, we uniformly sample an instance from the unobserved interactions $\bm{v}/\bm{v}_{+}^i$ (i.e., unpurchased items of the user) as the negative instance.
It should be noted that a nonuniform sampling strategy might further improve the performance, and we leave the exploration as a future work.
%negative instances $\bm{v}_{-}^i = \bm{v}/\bm{v}_{+}^i$. 

In this equation, we maximize the likelihood of our predicted results with the first two terms, and regularize all of the model parameters to avoid over fitting with the last term. In the training phase, we learn the parameters based on stochastic gradient descent (SGD) optimization. Once the model is optimized, we are not only able to generate a personalized recommendation list for a target user according to the predicted scores (i.e. $\hat y_{ij}$), but also can highlight particular regions of the corresponding product image as the visual explanations according to the attention weights (i.e. $\alpha_{i,j,k}$), which will be explained in the following parts of the paper.

%\subsection{The advanced model guided by textual features}
\subsection{The Review-enhanced Model}
\begin{figure}[t!]
\centering
\setlength{\fboxrule}{0.pt}
\setlength{\fboxsep}{0.pt}
\fbox{
\includegraphics[width=1.\linewidth,]{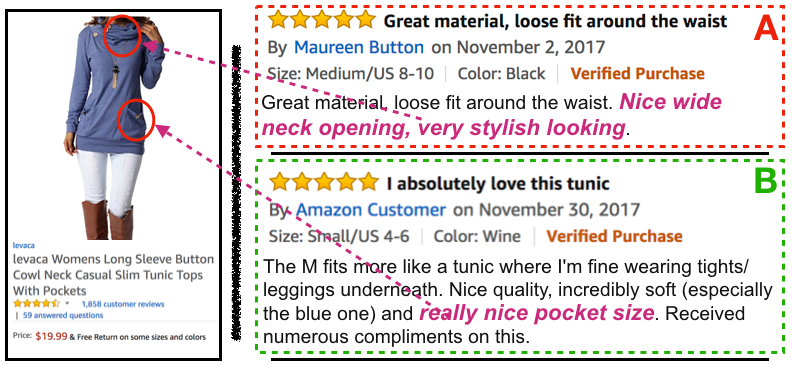}
}
\caption{An example of user reviews on Amazon. The pink italic fonts reveal user preferences that can be aligned with the corresponding visual features on the product image.}
\label{ex}
\vspace{-10pt}
\end{figure}

We have introduced the basic model for visually explainable recommendation based only on the visual images and implicit feedbacks. In many practical systems such as e-commerce, users usually express their opinions in the form of textual reviews.
Compared with pure implicit feedback, the textual review signals can be very helpful in our task because: (1) They provide explicit information that reveals user preferences. See the example in Figure~\ref{ex}, although user A and B bought the same top (i.e., both have a positive implicit feedback on the top), the features that they care about can be very different according to the posted reviews. User A cares more about the fitting and the neck opening, while user B is more interested in its quality and the pocket. Therefore, incorporating user reviews in the modeling process can help us to capture more comprehensive user preferences, which may lead to improved recommendation performance; (2) People may directly express their opinions on the visual features through their reviews. In the above example, user A expressed her opinion on the neck of the top in the image by ``... Nice wide neck opening, very stylish looking ...''. As a result, the textual reviews may exhibit as important signals to identify user preferences in the product image, which may help us to highlight more accurate visual regions tailed for different users, and further generate better visually explainable recommendations.

Motivated by these intuitions, we introduce user reviews as a weak supervision signal into our base model, and proposed review-enhanced visually explainable recommendation (Re-VECF for short) to further enhance the performance as well as the interpretability of the recommendations.

\subsubsection{\textbf{Textual feature modeling based on gated recurrent unit (GRU) network}}
Suppose $\bm{w_{ij}} = \{{w_{ij}^1}, {w_{ij}^2}, ..., {w_{ij}^{l_{ij}}}\}$ ($i\in \bm{u}, j \in \bm{v}$) is the review of user $i$ on item $j$, where ${w_{ij}^t}~(t\in \{1,2,...,l_{ij}\})$ is the word at time step $t$, and $l_{ij}$ is the length of the review. To model such textual features, we make use of recurrent neural networks (RNN)~\cite{mikolov2010recurrent}, which has been successfully applied to a number of language modeling tasks such as machine translation~\cite{bahdanau2014neural}, image captioning~\cite{vinyals2015show}, and video classification~\cite{yue2015beyond}. Specifically, we adopt the gated recurrent unit (GRU) network~\cite{cho2014properties} because we find it more computationally efficient in our task than the other RNN variations, such as the long-short term memory (LSTM) network~\cite{hochreiter1997long}. 

In a standard GRU network, the prediction of the current word is conditioned on the previous hidden states as well as the previously generated words. Computations at each time step are:
\begin{align}
&\bm{z}_{t} = \sigma (\bm{W}^z_{g}\bm{E}\bm{c}_{ij}^{t-1} + \bm{U}^z_{g}\bm{h}_{t-1} + \bm{b}^{z}_g)\\
&\bm{r}_{t} = \sigma (\bm{W}^r_{g}\bm{E}\bm{c}_{ij}^{t-1} + \bm{U}^r_{g}\bm{h}_{t-1} + \bm{b}^{r}_g)\\
&\bm{\tilde{h}}_{t} = \text{tanh}(\bm{W}^h_{g}\bm{E}\bm{c}_{ij}^{t-1} + \bm{U}^h_{g}(\bm{r}_{t}\circ \bm{h}_{t-1}) + \bm{b}^{h}_g)\\
&\bm{h}_{t} = \bm{z}_{t}\circ \bm{h}_{t-1} + (1-\bm{z}_{t})\circ \bm{\tilde{h}}_{t}
\end{align}
where $\bm{r}_{t}$ is the reset gate and $\bm{z}_{t}$ is the update gate.
$\bm{E} \in R^{O\times N^w}$ is the word embedding matrix, where $N^w$ is the vocabulary size, and $\bm{c}_{ij}^{t-1}\in R^{N^w}$ is the one-hot format of the input word $w_{ij}^{t-1}$. 
$\bm{h}_{t} \in R^Z$ is the hidden state, while $\bm{W}^z_{g}, \bm{W}^r_{g}, \bm{W}^h_{g} \in R^{Z\times O}$ and $\bm{U}^z_{g}, \bm{U}^r_{g}, \bm{U}^h_{g} \in R^{Z\times Z}$ are the parameter matrices, and $\bm{b}^z_g, \bm{b}^r_g, \bm{b}^h_g \in R^{Z}$ are the bias vectors.
Finally, $\circ$ denotes the element-wise multiplication function, and $\sigma(\cdot)$ and $\text{tanh}(\cdot)$ are the sigmoid and hyperbolic tangent activation functions, respectively.

\subsubsection{\textbf{Review-enhanced visually explainable collaborative filtering (Re-VECF)}}
The architecture of our model is shown in Figure~\ref{re-vec-nn}. To provide available signals to discover visual preferences, the prediction of each word is not only influenced by the previous word and the hidden state as in standard GRU, but also linked with the image regional features. More formally, the merged regional feature $\bm{IMAGE}_{j} = \sum_{k=1}^h \alpha_{i, j, k} \cdot \bm{f}_j^k$ is added into the reset and update gates to influence the review generation process:
\begin{align}
\label{GRU-0}
&\bm{z}_{t} = \sigma (\bm{W}^z_{g}\bm{E}\bm{c}_{ij}^{t-1} + \bm{U}^z_{g}\bm{h}_{t-1} + \bm{V}^z_{g}\bm{IMAGE}_j + \bm{b}^{z}_g)\\\label{GRU-1}
&\bm{r}_{t} = \sigma (\bm{W}^r_{g}\bm{E}\bm{c}_{ij}^{t-1} + \bm{U}^r_{g}\bm{h}_{t-1} + \bm{V}^z_{g}\bm{IMAGE}_j + \bm{b}^{r}_g)\\\label{GRU-2}
&\bm{\tilde{h}}_{t} = \text{tanh}(\bm{W}^h_{g}\bm{E}\bm{c}_{ij}^{t-1} + \bm{U}^h_{g}(\bm{r}_{t}\circ \bm{h}_{t-1}) + \bm{b}^{h}_g)\\\label{GRU-3}
&\bm{h}_{t} = \bm{z}_{t}\circ \bm{h}_{t-1} + (1-\bm{z}_{t})\circ \bm{\tilde{h}}_{t}
\end{align}
where $\bm{V}^z_{g} \in R^{Z\times D}$ is the parameter matrix for the visual features.
With the help of such a design, the user preference information embedded in the textual features can be leveraged to guide the learning of visual attentions (i.e. $\alpha_{i, j, k}$) through back propagation signals, denoted as the red dotted line in Figure~\ref{re-vec-nn}.

For simplicity, we abbreviate the computations from~Eq.\eqref{GRU-0} to \eqref{GRU-3} as:
\begin{align}
\bm{h}_{t} = \text{GRU}(\bm{h}_{t-1}, w_{ij}^{t-1}, \bm{IMAGE}_j)
\end{align}
and the word at time step $t$ can be predicted by:
\begin{align}
p(w_{ij}^t|\bm{w}_{ij}^{1:t-1},\bm{IMAGE}_j) = \text{SOFTMAX}(\bm{h}_{t}),~~~t\in \{2,3,...l_{ij}\}
\end{align}
where $\text{SOFTMAX}(\cdot)$ is an $N^{w}$-way softmax operation, and $\bm{w}_{ij}^{1:t-1} = \{w_{ij}^{t-1},w_{ij}^{t-2},...,w_{ij}^{1}\}$ is the set of all previous words before iteration $t$.
Note that when $t$=1, the sequence model has no input information, and we thus only use $\bm{IMAGE}_j$ to derive $\bm{h}_{1}$:
\begin{align}
&p(w_{ij}^1|\bm{IMAGE}_j) = \text{SOFTMAX}(\bm{h}_{1})\\
&\bm{h}_{1} = \text{GRU}(\bm{IMAGE}_j)
\end{align}

In real-world scenarios, textual reviews may associate with user preferences or item features that are not reflected in the product image. 
See the example in Figure~\ref{ex} again, in the review of user B \textit{``... Nice quality ...''}, the feature \textit{quality} can hardly be expressed in an image. 
Inspired by this intuition and to make our model more robust, we include user/item embeddings into the word prediction process to capture image-independent factors. Formally, we introduce a \textit{gate function $\beta$} to model whether the current word is generated from the image features or the user/item embeddings in a soft manner, and the above computations are thus further improved as:  
\begin{align}
&p(w_{ij}^1|\bm{z}^0_{ij}) = \text{SOFTMAX}(\bm{h}_{1})\\
&\bm{h}_{1} = \text{GRU}(\bm{z}^0_{ij})\\
&\bm{z}^0_{ij} = g\Big(\frac{1}{2}\big[(\bm{W}^{u}_{c})^T\cdot \bm{p}_{i} + (\bm{W}^{i}_{c})^T\cdot \bm{q}_{j} + (\bm{W}_{c}^{img})^T \nonumber \cdot \bm{IMAGE}_j\big]+\bm{b}_{c}\Big)
\end{align}
for the initial state, and for the subsequent states:
\begin{align}
p(&w_{ij}^t |\bm{w}_{ij}^{1:t-1},\bm{z}^{t-1}_{ij}) = \text{SOFTMAX}(\bm{h}_{t}),~~~t\in \{2,3,...l_{ij}\}\\
\bm{h}_{t} &=  \text{GRU}(\bm{h}_{t-1}, w_{ij}^{t-1}, \bm{z}^{t-1}_{ij})\\
\bm{z}^t_{ij} &=  g\Big(\big[(\bm{W}^{u}_{c})^T\cdot \bm{p}_{i} + (\bm{W}^{i}_{c})^T\cdot \bm{q}_{j}\big]\cdot \beta\big((\bm{w}^{h}_{c})^T\cdot\bm{h}_{t}\big)\nonumber\\
&+ \big[(\bm{W}_{c}^{img})^T\cdot \bm{IMAGE}_j\big]\cdot\big(1-\beta((\bm{w}^{h}_{c})^T\cdot\bm{h}_{t})\big) + \bm{b}_{c}\Big)
\end{align}
where $\bm{z}^t_{ij}$ is a context vector used to influence the review generation process. $\beta(\cdot)$ is the sigmoid function used to weigh the image and user/item embeddings. $g(x) = \max(0, x)$ is the Rectified Linear Unit (ReLU)~\cite{maas2013rectifier} active function, and $\bm{W}^{u}_{c},\bm{W}^{i}_{c},\bm{W}^{img}_{c},\bm{w}^{h}_{c},\bm{b}_{c}$ are the model parameters. In the computation of $\bm{z^0_{ij}}$, we initialize $\beta=\frac{1}{2}$ so that the image embedding and user/item embeddings are equally important.

At last, by simultaneously predicting user implicit feedback and textual reviews, our final objective function to be maximized is:
\begin{equation}
\begin{aligned}\label{eq:l2}
l_{2} &= \delta \sum_{(i,j)}\sum_{t=1}^{l_{ij}} \log~p(w_{ij}^{t}|\bm{w}_{ij}^{1:t-1},\bm{z}^{t-1}_{ij}) \\
&+(1-\delta) \left(\sum_{i\in \bm{u}}\sum_{j \in \bm{v}_{+}^i}\log \hat y_{ij} + \sum_{i\in \bm{u}}\sum_{j \in \bm{v}/\bm{v}_{+}^i} \log(1-\hat y_{ij}) \right)- \lambda ||\Theta||^2_F
\end{aligned}
\end{equation}

In this equation, we are formulating our problem into a multi-task learning framework. By jointly capturing the preferences from user implicit feedbacks and textual reviews, we aim to achieve both recommendation accuracy and reasonable high-quality visual explanations.

\section{Experiments}
In this section, we evaluate our proposed models focusing on the following three key research questions: \\

\noindent
$\textbf{RQ 1: }$ Performance of our models for \textit{Top-N recommendation}.\\
\noindent
$\textbf{RQ 2: }$ Performance of our models for \textit{review prediction}.\\  
\noindent
$\textbf{RQ 3: }$ Performance of our models for \textit{providing visual explanations}.\\

As mentioned before, our final model \textbf{Re-VECF} can be seen as a multi-task learning framework. The first two questions are designed to evaluate each subtask one by one, while the third question aims to study the visual explanations provided by our models.
We begin by introducing the experimental setup, and then report and analyze the experimental results to answer these research questions.

\subsection{Experimental Setup}\label{set}

\noindent
\textbf{Datasets}. We conduct our experiments on the Amazon e-commerce dataset\footnote{http://jmcauley.ucsd.edu/data/amazon/} ~\cite{he2016ups,mcauley2015image}.
This dataset contains user-product purchasing behaviors as well as product images and textual reviews from Amazon spanning May 1996 - July 2014. 
We evaluate our models on the categories of Clothing, Shoes and Jewelry/Men and Clothing, Shoes and Jewelry/Women with the statistics shown in Table~\ref{tb-dataset}.

\begin{table}[!h]
\centering
\caption{Statistics of the datasets.}
\vspace{-10pt}
\begin{tabular}{p{1cm}<{\centering}|p{0.8cm}<{\centering}|p{0.8cm}<{\centering}|p{1.6cm}<{\centering}|p{1cm}<{\centering}|p{1cm}<{\centering}}
\hline\hline
       Datasets               &\#Users   &\#Items&\#Interactions &Density &\#Words \\ \hline
{Men}&643&2454&6359&0.403\%&21600\\\hline
{Women}&570&3346&7640&0.401\%&17614 \\\hline\hline
\end{tabular}
\label{tb-dataset}
\end{table}

\noindent
\textbf{Evaluation methods}. We use the following measures for evaluation on different tasks:

$\bullet$ \textbf{Precision (P), Recall (R) and $F_{1}$-score ($F_1$): } These measures aim to evaluate the quality of the recommendations~\cite{karypis2001evaluation}. In the context of recommendation system, Precision computes the percentage of correctly recommended items in a user's recommendation list, averaged across all testing users. Recall computes the percentage of purchased items that are really recommended in the list, and it is also averaged across all testing users. By considering both Precision and Recall, $F_{1}$-score computes the harmonic average between them, which is reported in our experiments as the final results. 

$\bullet$ \textbf{Hit-Ratio (HR): } Hit-ratio gives the percentage of users that can receive at least one correct recommendation, which has been widely used in previous work~\cite{karypis2001evaluation,xiang2010temporal}.

$\bullet$ \textbf{NDCG: } To assess if the items that a user has actually consumed are ranked in higher positions in the recommendation list, we use normalized discounted cumulative gain (NDCG) to evaluate ranking performance by taking the positions of the correct items into consideration~\cite{jarvelin2000ir}.

$\bullet$ \textbf{ROUGE: } ROUGE score~\cite{lin2004rouge} is a widely used metric for evaluating the quality of text generation. It computes the overlapping of n-grams between the generated text and the ground truth. In our model, take 2-grams for example, the predicted review $\bm{\hat s}_{ij}$ and the true review $\bm{s}_{ij}$ are first mapped into 2-gram sets $G(\bm{\hat s}_{ij}) = \big\{({\hat s_{ij}}^1, {\hat s_{ij}}^2),({\hat s_{ij}}^2, {\hat s_{ij}}^3),...({\hat s_{ij}}^{l_{ij}-1}, {\hat s_{ij}}^{l_{ij}})\big\}$ and $G(\bm{s}_{ij}) = \big\{(s_{ij}^1, s_{ij}^2),(s_{ij}^2, s_{ij}^3),\\...(s_{ij}^{o_{ij}-1}, s_{ij}^{o_{ij}})\big\}$, respectively. Then the Precision (ROUGE-2-P), Recall (ROUGE-2-R) and $F_{1}$-score (ROUGE-2-$F_{1}$) are computed as:
\begin{equation}
\begin{aligned}
\text{ROUGE-2-P} &= \frac{|G(\bm{s}_{ij}) \cap G(\bm{\hat s}_{ij})|}{|G(\bm{\hat s}_{ij})|}\\
\text{ROUGE-2-R} &= \frac{|G(\bm{s}_{ij}) \cap G(\bm{\hat s}_{ij})|}{|G(\bm{s}_{ij})|}\\
\text{ROUGE-2-$F_{1}$} &= \frac{2\times \text{ROUGE-2-P} \times \text{ROUGE-2-R}}{\text{ROUGE-2-P}+\text{ROUGE-2-R}}
\end{aligned}
\end{equation}
In this work, we report ROUGE score under 1-gram and 2-gram settings, referred to as ROUGE-1 and ROUGE-2, respectively.\\

\begin{table}[!t]
\caption{Summary of the models in our experiments on the three tasks, respectively, which compares the specific information used in each model and the depth of the models.}
\vspace{-5pt}
\centering
\setlength{\tabcolsep}{1pt}
\begin{tabular}
{p{1.8cm}<{\centering}p{1.8cm}<{\centering}p{1.8cm}<{\centering}p{2.0cm}<{\centering}} \hline\hline
Model&Reference&Information&Depth\\\hline\hline
\multicolumn{2}{c}{\emph{\textbf{Top-N recommendation}}}&\multicolumn{2}{c}{\emph{\textbf{Measures:~~$F_1$, HR and NDCG}}}\\
\cline{2-3}
BPR&\cite{rendle2009bpr}&-&shallow model\\
VBPR&\cite{he2016vbpr}&image&shallow model\\
HFT&\cite{mcauley2013hidden}&text&shallow model\\
NRT&\cite{li2017neural}&text&deep model\\
JRL&\cite{zhang2017joint}&image+text&deep model\\
VECF&section 4.1&image&deep model\\
Re-CF&section 4.2&text&deep model\\
Re-VECF&section 4.2&image+text&deep model\\\hline
\multicolumn{2}{c}{\emph{\textbf{Review prediction}}}&\multicolumn{2}{c}{\emph{\textbf{Measures:~~ROUGE}}}\\
\cline{2-3}
NRT&\cite{li2017neural}&text&deep model\\
Re-CF&section 4.2&text&deep model\\
Re-VECF&section 4.2&image+text&deep model\\\hline
\multicolumn{2}{c}{\emph{\textbf{Visual explanation}}}&\multicolumn{2}{c}{\emph{\textbf{Measures:~~$F_1$ and NDCG}}}\\
\cline{2-3}
VECF&section 4.1&image&deep model\\
Re-VECF&section 4.2&image+text&deep model\\\hline
\hline
\end{tabular}\label{tab:eval-m}
\vspace{-10pt}
\end{table}

\noindent
\textbf{Baselines}. We adopt the following representative and state-of-the-art methods as baselines for performance comparison:
%$\bullet$ \textbf{MostPopular (MP):} This is a non-personalized static method, where the most frequently purchased items are ranked in descending order of frequency to make recommendations.

$\bullet$ \textbf{BPR: } The bayesian personalized ranking~\cite{rendle2009bpr} model is a popular method for top-N recommendation. We adopt matrix factorization as the prediction component for BPR.

$\bullet$ \textbf{VBPR: } The visual bayesian personalized ranking~\cite{he2016vbpr} model is a state-of-the-art method for recommendation leveraging product visual images.

$\bullet$ \textbf{HFT: } The hidden factors and topics model~\cite{mcauley2013hidden} is a well known recommendation method leveraging user textual reviews.

$\bullet$ \textbf{NRT: } The neural rating regression model~\cite{li2017neural} is a state-of-the-art neural recommender which can generate user textual reviews.

$\bullet$ \textbf{JRL: } The joint representation learning model~\cite{zhang2017joint} is a state-of-the-art neural recommender, which can leverage multi-model side information for Top-N recommendation. 

$\bullet$ \textbf{VECF: } This is the base model proposed in section 4.1, it integrates visual attention mechanism with collaborative filtering under the supervision of user implicit feedbacks. 

$\bullet$ \textbf{Re-CF: } This is a variation of our final model in section 4.2. We remove the image feature from the input, and derive an image-free collaborative filtering model supervised by the user implicit feedback and textual reviews. 

And our final model for visually explainable recommendation is denoted as \textbf{Re-VECF}. For easy understanding, we summarize the similarities and differences of all the models in our experiments on different tasks in Table~\ref{tab:eval-m}.\\
%Finally, our unified model for visually explainable recommendation introduced (section 4.2) is denoted as \textbf{Re-VisExpCF} in the following experiments.

\noindent
\textbf{Parameter settings}. We initialize all the optimization parameters according to a uniform distribution in the range of $(0,1)$, and update them by conducting stochastic gradient descent (SGD). We determine the learning rate and the tuning parameter $\delta$ (in Eq.\eqref{eq:l2}) in the range of $\{1,0.1,0.01^*,0.001\}$ and $\{0.1, 0.2^*, 0.3, ..., 0.8, 0.9\}$, respectively, where $^*$ indicates the final values used in our experiments. 
The dimension of user/item embedding $K$ is tuned in the range of $\{10, 20, 30, ..., 90, 100\}$. 
For each dataset, we first tokenize all the user reviews by the Stanford Core NLP tool\footnote{https://stanfordnlp.github.io/CoreNLP/}, and then build the lexicon by retaining all the tokens for GRU training.
The word embeddings are pre-trained based on the Skip-gram model\footnote{http://mccormickml.com/2016/04/19/word2vec-tutorial-the-skip-gram-model/}, and the embedding size is set as 64.
For the baselines, we determine the optimal parameter settings by grid search, and the models designed for rating prediction (i.e. HFT and NRT) are learned by optimizing the pairwise ranking loss of BPR to model user implicit feedback. 
%For JRL, we only include textual and visual features in the modeling process because we are using implicit feedbacks.
When conducting experiments, 70\% items of each user are leveraged for training, while the remaining are used for testing. We generate top-5 recommendation list ($n=5$) for each user in the test dataset.

\subsection{Top-N Recommendation (\textbf{RQ1})}
In this section, we evaluate our models for the task of Top-N recommendation. 
Specifically, we first compare our models Re-CF, VECF and Re-VECF with the previously proposed methods (i.e. BPR, VBPR, HFT, NRT, and JRL), and then we study the influence of embedding dimension $K$ on the recommendation results. \\

\noindent
\textbf{Model comparison.}
Table~\ref{tab:result} shows the performance of different methods on $F_{1}$, HR and NDCG, we can see that,

$\bullet$ VBPR, HFT, NRT, Re-CF and VECF can achieve better performance than BPR. On considering that the key difference between BPR and VBPR/HFT/NRT/Re-CF/VECF is that the latter models integrate either visual or textual features into their modeling process, this observation verifies that side information -- such as user reviews or product images -- can help to improve the performance in real-world systems. Furthermore, by incorporating both visual and textual features together, JRL obtains the best performance among the baselines.

$\bullet$ NRT and Re-CF can achieve better performance than HFT on both Men and Women datasets. This is within expectation because 1) the multiple non-linear layers in NRT can be more expressive in terms of user preference modeling compared with the inner product operation in HFT, and 2) NRT and Re-CF can better capture word sequential information than HFT in review sentences. This allows NRT and Re-CF to achieve better precision in textual feature and user profile modeling, which further improves the recommendation performance.

\begin{table}[!t]
\caption{Summary of the performance for baselines and our models. The first block shows the baseline performances, where starred numbers are the best baseline results; the second block shows the results of Re-VECF and its variations. Bolded numbers are the best performance of each column, and all numbers in the table are percentage numbers with `\%' omitted. Improvements of our final model Re-VECF from the best baseline are significant at $p=0.01$ with paired $t$-test.}
\setlength{\tabcolsep}{1.9pt}
\begin{tabular}
{c|ccc|ccc} \hline\hline
Dataset & \multicolumn{3}{c|}{Men} & \multicolumn{3}{c}{Women} \\\hline
Measure@5(\%)&$F_{1}$&HR&NDCG&$F_{1}$&HR&NDCG\\\hline
BPR   &1.209&3.901&0.740
      &0.897&3.342&0.611\\
HFT   &1.242&4.243&0.757
      &0.915&3.371&0.631\\
VBPR  &1.361&4.261&0.773
      &0.929&3.402&0.648\\
NRT   &1.399&4.469&0.802
      &0.952&3.527&0.674\\
JRL   &$1.424^*$&$4.703^*$&$0.813^*$
      &$0.967^*$&$3.542^*$&$0.686^*$\\\hline
Re-CF &1.370&4.364&0.781
      &0.937&3.451&0.651\\
VECF  &1.378&4.373&0.791
          &0.948&3.523&0.669\\
Re-VECF &\textbf{1.442}&\textbf{4.803}&\textbf{0.846}
            &\textbf{0.985}&\textbf{3.587}&\textbf{0.712}\\\hline\hline
\end{tabular}\label{tab:result}
\vspace{-15pt}
\end{table}

$\bullet$ VECF performs better than Re-CF, and VBPR performs better than HFT. This observation highlights the importance of visual features in personalized recommendation, and it is consistent with the intuition that customers are largely influenced by the product images when making purchasing decisions online, so that images contain rich information about users' personalized preferneces.

$\bullet$ The performance of Re-CF and VECF fail to surpass JRL. This observation is not surprising because JRL takes both review and image features for user/item profiling, while Re-CF and VECF takes only one of the information sources. Encouragingly, we find that our final Re-VECF model achieves better performance than JRL. 
This result indicates the effectiveness of our method for the Top-N recommendation task, and the main reason can be that when profiling visual features, JRL roughly takes a fixed vector to represent the whole product image, while in our model, the attention mechanism can provide us with the opportunity to discriminatingly focus on the image regions that are more important to the corresponding user, which eventually helps to better capture the user preferences and improve the recommendation performance.\\

\begin{figure}[t]
\centering
\setlength{\fboxrule}{0.pt}
\setlength{\fboxsep}{0.pt}
\fbox{
\includegraphics[width=0.9\linewidth]{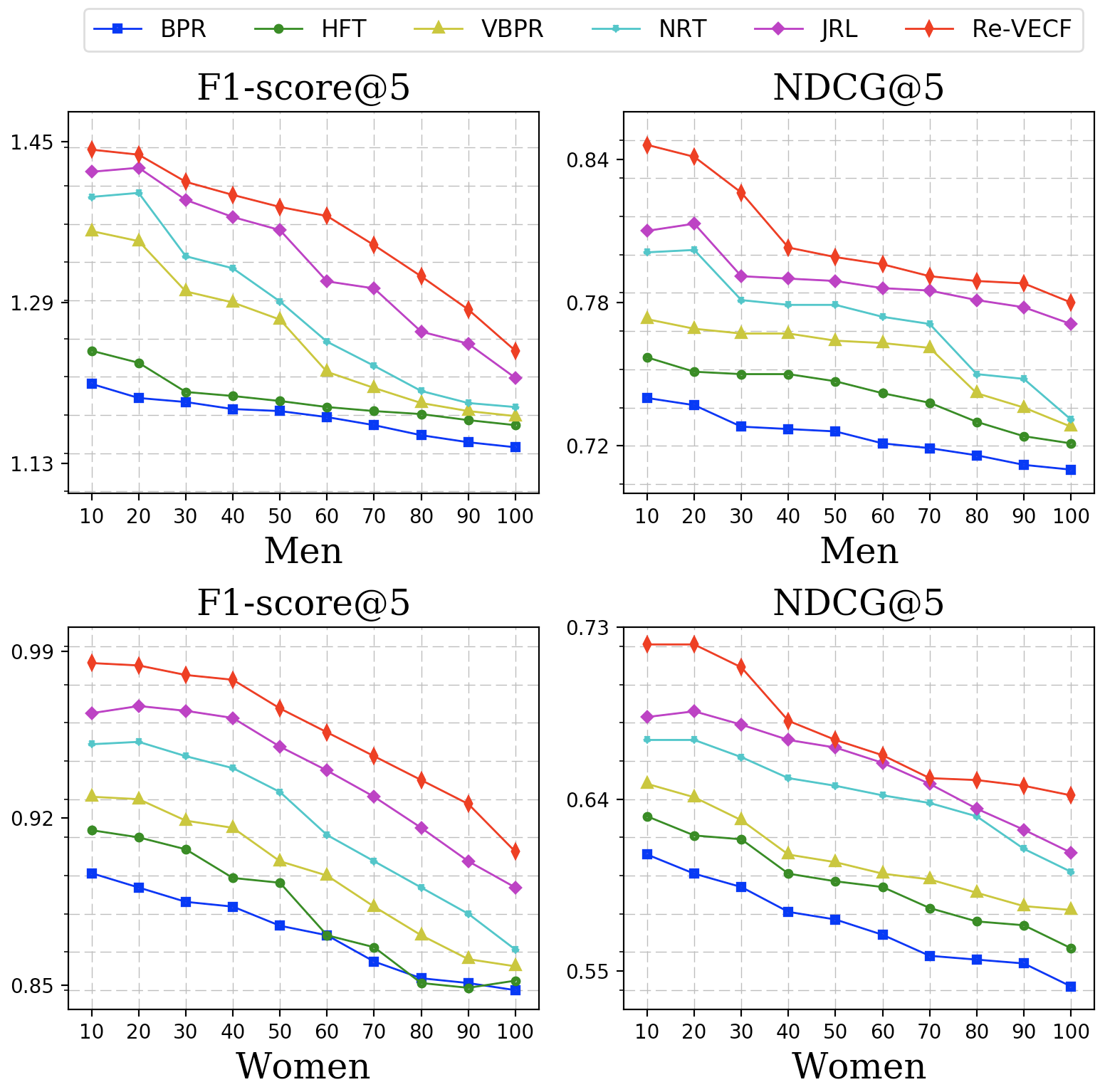}
}
\vspace{-12pt}
\caption{Performance of our models and baselines under difference choices of embedding dimension $K$.}
\label{para-K}
\vspace{-15pt}
\end{figure}

\noindent
\textbf{Influence of the embedding dimension $K$.}
In this section, we study how the embedding dimension influences the model performance. We set all other parameters according to section~\ref{set}, and observe the performance by tuning $K$ from 10 to 100 (using even larger $K$ values will decrease the performance). From the results on $F1$@5 and $NDCG$@5 shown in Figure~\ref{para-K}, we can see that on the Men dataset, all the models can reach their best performance when using some small dimensions, while using additional parameters does not help promoting the performance. Similar results can also be observed on the Women dataset. This observation suggests that while expressive power is increased, using too many latent factors may also increase the model complexity extremely and may lead to over-fitting, which can weaken the generalization ability of our models on the test dataset.

\subsection{Review Prediction (\textbf{RQ2})}
In this section, we evaluate the second subtask of our model -- review generation -- by comparing the predicted reviews with the truly posted ones.
Specifically, we first conduct quantitative evaluation on the models that can generate user reviews, and then we present intuitive analysis on the predicted reviews in a qualitative manner.

\noindent
\textbf{Quantitative evaluation.}
To begin with, we compare our final model Re-VECF with Re-CF and NRT, where Re-CF is the image-free version of Re-VECF. The model parameters follow the settings in section~\ref{set}, and we conduct this experiment on the Men dataset.
From the results on ROUGE-1 and ROUGE-2 shown in Table~\ref{tab:review-result}, we can see that 
Re-VECF achieves significantly better performance than Re-CF and NRT on all the metrics. 
This is as expected because the product aspects that users comment in textual reviews may be directly aligned with the product images, thus including visual features in the modeling process can effectively capture such signals and make accurate predictions.

\noindent
\textbf{Qualitative analysis.}
For the purpose of providing more intuitions, we also list several examples of the generated user reviews in Table~\ref{tab:review-generation}.
We can see that with the help of gated recurrent units (GRU) for natural language generation, the linguistic quality of the generated reviews from all the models are reasonably good. 
More encouragingly, Re-VECF can generate words that directly describe (part of) the product image, and some of the words are also mentioned in the true review. As in the boxed areas on the images for example, Re-VECF can generate very explicit descriptive words such as \textit{sleeve} and \textit{buckle} by automatically aligning the information between image and text, while Re-CF and NRT only output some very general-purpose expressions.
This observation is in line with the quantitative results mentioned above, and it further verifies that by including visual features in the modeling process, Re-VECF has the ability to learn the relationships between visual- and textual- features, and thus to generate descriptive textual expressions for the recommendations and visual images.

\begin{table}[!t]
\centering
\caption{Performance comparison between Re-CF, NRT and Re-VECF on the task of review prediction. Improvements of Re-VECF from baselines are significant at $p=0.01$ level.}
\vspace{-5pt}
\setlength{\tabcolsep}{1pt}
\begin{tabular}{p{1.5cm}<{\centering}|p{1cm}<{\centering}|p{1cm}<{\centering}|p{1cm}<{\centering}|p{1cm}<{\centering}|p{1cm}<{\centering}|p{1cm}<{\centering}}
\hline\hline
\multirow{2}{*}{Method}& \multicolumn{3}{c|}{ROUGE-1} & \multicolumn{3}{c}{ROUGE-2} \\
\cline{2-7}
&P(\%)&R(\%)&$F_{1}(\%)$&P(\%)&R(\%)&$F_{1}(\%)$\\\hline
Re-CF&16.15&37.98&19.11&1.11&3.73&1.61\\\hline
NRT&18.67&41.28&21.77&1.42&4.12&2.01\\\hline
Re-VECF&22.01&48.36&27.49&1.68&4.78&2.32\\\hline\hline
\end{tabular}\label{tab:review-result}
\end{table}

\begin{table}[!t]
\centering
\small
\caption{Examples of the generated reviews compared with the true reviews. The bolded italic words (e.g., sleeve) mean that the word generated by Re-VECF was also mentioned in the true review, and the word is aligned to the boxed area of the image learned by the attention mechanism in our model.}
\vspace{-5pt}
\setlength{\tabcolsep}{1pt}
\begin{tabular}{| >{\centering\vspace{1mm}}m{1.2cm}<{\vspace{1mm}} | >{\centering\vspace{1mm}}m{2.5cm}<{\vspace{1mm}} | >{\centering\vspace{1mm}}m{1.6cm}<{\vspace{1mm}} | >{\centering\vspace{1mm}}m{1.2cm}<{\vspace{1mm}} | >{\centering\vspace{1mm}}m{1.3cm}<{\vspace{1mm}}|}
\hline
Image & True Review & Re-VECF & Re-CF & NRT\tabularnewline\hline\hline
\parbox[c]{1.5cm}{\includegraphics*[scale=0.5]{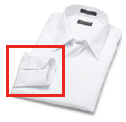}} & It's an excellent poplin solid color long \textit{\textbf{sleeved}} shirt & Much like the \textit{\textbf{sleeve}} & Not bad for the price & Very good choice\tabularnewline\hline
\parbox[c]{1.5cm}{\includegraphics*[scale=0.5]{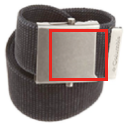}} & Very \textit{\textbf{good-looking}} sturdy belt with a good ribbed weave and strong \textit{\textbf{buckle}} & I like this \textit{\textbf{good looking}} \textit{\textbf{buckle}} & Great for the price & Makes a great price\tabularnewline\hline
\end{tabular}\label{tab:review-generation}
\vspace{-10pt}
\end{table}

\iffalse%
\begin{figure}[!t]
\centering
\setlength{\fboxrule}{0.pt}
\setlength{\fboxsep}{0.pt}
\fbox{
\includegraphics[width=0.95\linewidth]{fig/review-gen.png}
}
\caption{Examples of the generated reviews. The true reviews are denoted in black, and the predicted reviews from the Re-CF, NRT and Re-VisExpCF models are denoted in red, purple underlined and yellow italic fonts, respectively. The greened texts mean that the word \textit{sleeve} which was mentioned in the true review can also be generated by Re-VisExpCF, and the word is aligned to the corresponding visual parts of the image by our model. (best view in color)}
\label{review-generated}
\end{figure}
\fi%

\iffalse%
\begin{figure*}[!t]
\centering
\setlength{\fboxrule}{0.pt}
\setlength{\fboxsep}{0.pt}
\fbox{
\includegraphics[width=0.95\linewidth]{fig/visual-exp.png}
}
\caption{Examples of the visual explanations. 
Each row represents a user. 
The current item column lists the products to be evaluated in the test dataset.
We list two most similar purchased products to the current item in the historical records column, and the review column shows the comments of the current items from the corresponding users.
We compare the highlighted regions provided by VisExpCF and Re-VisExpCF for the same item. The right most two columns show the result from VisExpCF and Re-VisExpCF, respectively, and we label the words related to the visual explanations in yellow.}
\label{visual-exp}
\end{figure*}
\fi%

\begin{table*}[!t]
\centering
\caption{Examples of the visual explanations, where each row represents the target item of a user. The first column lists the image of the target item, and in the second column we list two most similar products to the target item that the user purchased before. The third column shows the user's review on the target item, and in the last two columns, we compare the highlighted regions provided by VECF and Re-VECF for the target item. In the review column, we use bolded italic to highlight the part of user review that our generated visual explanations correspond to.}
\vspace{-5pt}
\setlength{\tabcolsep}{2pt}

\begin{tabular}{| >{\centering\vspace{0mm}}m{0.5cm}<{\vspace{0mm}} | >{\centering\vspace{0mm}}m{1.8cm}<{\vspace{0mm}} | >{\centering\vspace{0mm}}m{2.5cm}<{\vspace{0mm}} | >{\centering\vspace{0mm}}m{8cm}<{\vspace{0mm}} | >{\centering\vspace{0mm}}m{1.5cm}<{\vspace{0mm}} | >{\centering\vspace{0mm}}m{1.5cm}<{\vspace{0mm}}|}
\hline
\multirow{2}{*}{\#} & \multirow{2}{*}{Target Item} & \multirow{2}{*}{Historical Records} & \multirow{2}{*}{Textual Review} & \multicolumn{2}{c|}{Visual Explanation}\tabularnewline
\cline{5-6}
& & & & VECF & Re-VECF\tabularnewline\hline\hline
 \end{tabular}
 
 \begin{tabular}{| >{\centering\vspace{0.5mm}}m{0.5cm}<{\vspace{0.5mm}} | >{\centering\vspace{0.5mm}}m{1.8cm}<{\vspace{0.5mm}} | >{\centering\vspace{0.5mm}}m{2.5cm}<{\vspace{0.5mm}} | >{\centering\vspace{0.5mm}}m{8cm}<{\vspace{0.5mm}} | >{\centering\vspace{0.5mm}}m{1.5cm}<{\vspace{0.5mm}} | >{\centering\vspace{0.5mm}}m{1.5cm}<{\vspace{0.5mm}}|}
1 & \parbox[c]{1cm}{\includegraphics*[height=0.9cm, width=0.9cm]{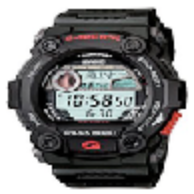}} & \parbox[c]{2cm}{\includegraphics*[height=0.9cm, width=2cm]{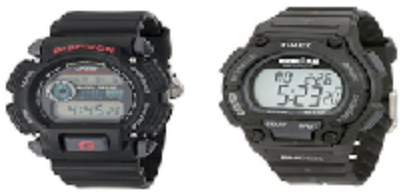}} & this is a large watch... nearly as large as my suunto but due to \textit{\textbf{its articulated strap it fits on the wrist very well.}} & \parbox[c]{1cm}{\includegraphics*[height=0.9cm, width=0.9cm]{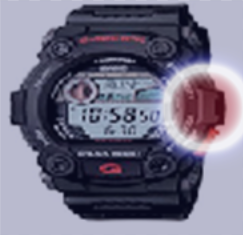}} & \parbox[c]{1cm}{\includegraphics*[height=0.9cm, width=0.9cm]{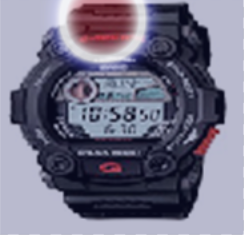}}\tabularnewline\hline

2 & \parbox[c]{1cm}{\includegraphics*[height=0.9cm, width=0.9cm]{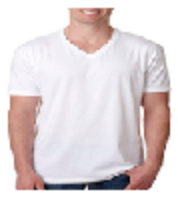}} & \parbox[c]{2cm}{\includegraphics*[height=0.9cm, width=2cm]{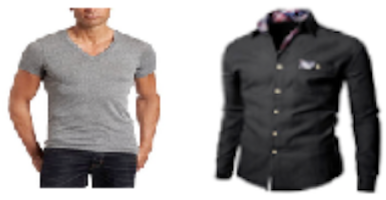}} & \textit{\textbf{this is a really comfortable v-neck. i found that the size and location of the v are just right for me.}} i'm 5'8 \& \#34, but 200 lbs ( and dropping :) ) & \parbox[c]{1cm}{\includegraphics*[height=0.9cm, width=0.9cm]{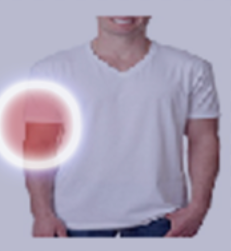}} & \parbox[c]{1cm}{\includegraphics*[height=0.9cm, width=0.9cm]{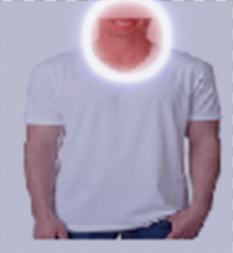}}\tabularnewline\hline

3 & \parbox[c]{1cm}{\includegraphics*[height=0.9cm, width=0.9cm]{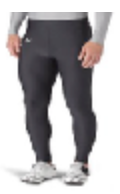}} & \parbox[c]{2cm}{\includegraphics*[height=0.9cm, width=2cm]{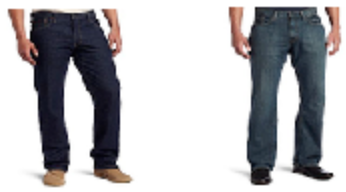}} & \textit{\textbf{Great leggings. perfect for fly fishing or hunting or running.}} just perfect anytime you are cold! & \parbox[c]{1cm}{\includegraphics*[height=0.9cm, width=0.9cm]{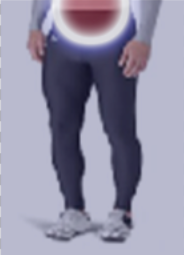}} & \parbox[c]{1cm}{\includegraphics*[height=0.9cm, width=0.9cm]{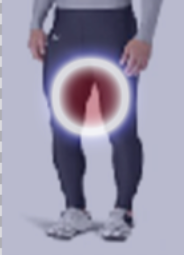}}\tabularnewline\hline

4 & \parbox[c]{1cm}{\includegraphics*[height=0.9cm, width=0.9cm]{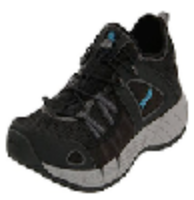}} & \parbox[c]{2cm}{\includegraphics*[height=0.9cm, width=2cm]{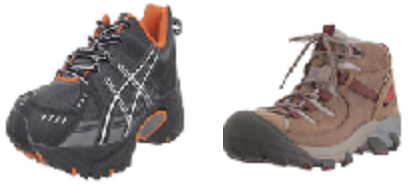}} & The socks on the shoes are a perfect fit for me. \textit{\textbf{first time with a shoe with the speed laces and i like them a lot}} & \parbox[c]{1cm}{\includegraphics*[height=0.9cm, width=0.9cm]{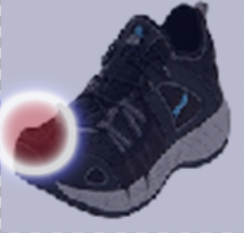}} & \parbox[c]{1cm}{\includegraphics*[height=0.9cm, width=0.9cm]{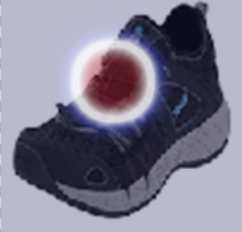}}\tabularnewline\hline

5 & \parbox[c]{1cm}{\includegraphics*[height=0.9cm, width=0.9cm]{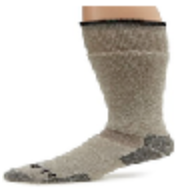}} & \parbox[c]{2cm}{\includegraphics*[height=0.9cm, width=2cm]{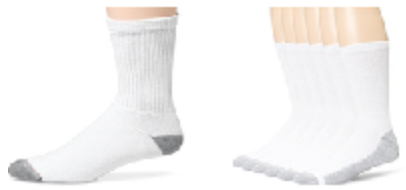}} & Really like these socks! they are really thick woolen socks and are good for cold days. \textit{\textbf{they cover a good portion of your feet as they go a little (halfway) above the calf muscle area}}. & \parbox[c]{1cm}{\includegraphics*[height=0.9cm, width=0.9cm]{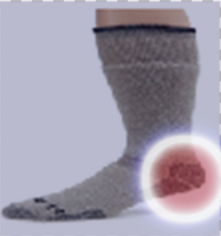}} & \parbox[c]{1cm}{\includegraphics*[height=0.9cm, width=0.9cm]{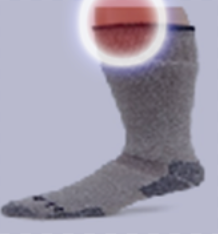}}\tabularnewline\hline

6 & \parbox[c]{1cm}{\includegraphics*[height=0.9cm, width=0.9cm]{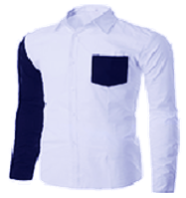}} & \parbox[c]{2cm}{\includegraphics*[height=0.9cm, width=2cm]{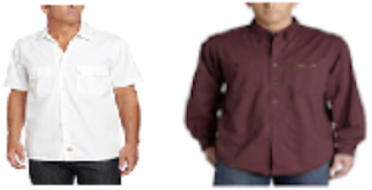}} & \textit{\textbf{I like the front pocket$\sim$!}} Very cool! & \parbox[c]{1cm}{\includegraphics*[height=0.9cm, width=0.9cm]{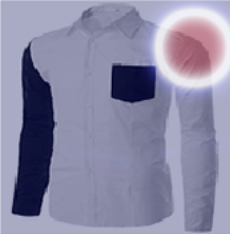}} & \parbox[c]{1cm}{\includegraphics*[height=0.9cm, width=0.9cm]{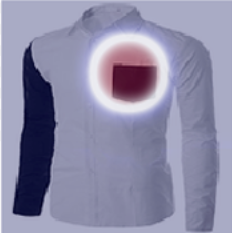}}\tabularnewline\hline
\end{tabular}\label{tab:visual-explanation}
\end{table*}

\subsection{Visual Explanation (\textbf{RQ3})}
In this section, we evaluate whether the visual explanations generated by our model are reasonable, i.e., whether the highlighted regions of the image learned by our model really reveal a user's potential interests on the recommended item. Similarly, we also conduct quantitative analysis first based on a dataset with collectively labeled ground-truth. Then, to provide better intuitions for the generated visual explanations, we present and analyze several examples learned by the model in a qualitative manner.

\textbf{Quantitative evaluation.}
To the best of our knowledge, this work is the first one on visually explainable recommendation, and there is no publicly available dataset with labeled ground-truth to evaluate whether the visual explanations (i.e., the visual highlights) generated by our model are reasonable or not. To tackle the problem, we build a collectively labeled dataset with Amazon MTurk by asking the workers to identify the image regions that may explain why a user bought a particular item, based on the user's previous purchase records and his/her review written on the target item.

More specifically, we still adopt the Amazon dataset, and retain the top-100 most active users (i.e., users with the most purchasing records) in the dataset. These users are provided to MTurk workers for labeling, so that the workers have sufficient historical information about a user to understand the user's personalized interests when labeling an image for the user.

For each of the 100 users, we randomly select one item that the user purchased before as the target item to label, and the image of the target item is equally divided into $5\times5=25$ square regions. A label task for a worker is to identify 5 out of the 25 regions that the worker believes are most relevant to the user's preference. For each label task, we provide the following two information sources to the worker for reference:
\begin{itemize}
\item Images and the corresponding reviews of the products that the user previously purchased.
\item The user's review on the target item to be labeled.
\end{itemize}

In a label task, a worker is first required to read the image-review pairs of the user's previously purchased products (around 10 pairs). After that, the worker will be shown the target image as well as the corresponding review, and be asked to identify 5 regions of the image. In this way, the worker can understand the user's personalized preference through the user reviews, and then identify the relevant image regions based on both the user preference and the user's review on the target item.

%In the test dataset, we only select one item for a user, and for each targeted user-item pair, we provide the subjects with: 
%\romannumeral1. all the images and the corresponding reviews of the user's previously interacted items, and \romannumeral2. the user's review for the current item. The subjects are required to label 5 out of 196 regions on the current item image. 

\begin{table}[!t]
\centering
\caption{Basic statistics of the labeled dataset.}
\vspace{-10pt}
\begin{tabular}{p{1cm}<{\centering}|p{0.8cm}<{\centering}|p{2.8cm}<{\centering}|p{2.5cm}<{\centering}}
\hline\hline
\#Users&\#Items&\#identified regions & \#regions/image\\ \hline
94&94&220&2.34\\\hline\hline
\end{tabular}
\vspace{-5pt}
\label{visual-dataset}
\end{table}

\begin{table}[!t]
\centering
\caption{Performance comparison between VECF and Re-VECF on visual explanation task by identifying top-5,10 relevant regions out of 196 candidate regions. Improvements of Re-VECF from VECF are significant at $p=0.01$ level.}
\vspace{-10pt}
\setlength{\tabcolsep}{1.9pt}
\begin{tabular}{p{1.8cm}<{\centering}|p{1.5cm}<{\centering}|p{1.5cm}<{\centering}|p{1.5cm}<{\centering}|p{1.5cm}<{\centering}}
\hline\hline
\multirow{2}{*}{Method}& \multicolumn{2}{c|}{Top-5} & \multicolumn{2}{c}{Top-10} \\
\cline{2-5}
&$F_{1}(\%)$&NDCG(\%)&$F_{1}(\%)$&NDCG(\%)\\\hline
Random &3.22&8.24&7.41&11.46\\\hline
VECF&6.70&17.37&10.38&16.40\\\hline
Re-VECF &8.35&20.53&12.99&19.95\\\hline\hline
\end{tabular}\label{tab:visual-result}
\vspace{-15pt}
\end{table}

Finally, each target item is labeled by two workers, and we only retain the common regions identified by both workers as the ground-truth, thus the final number of regions for an image may be less than 5. Some basic statistics of the labeled dataset are shown in Table~\ref{visual-dataset}. Note that the final number of users and items are less than 100 because there are 6 target items for which the workers have no commonly identified region.
%and if there are no overlapping labels between two people for a sample, it will be filtered out

For evaluation, we compare VECF and Re-VECF as no other models can provide visual explanations, and the model parameters follow the settings in section \ref{set}. Because both VECF and Re-VECF models work on $14\times14=196$ image features, we use each model to identify the top-5 and 10 regions out of the 196 candidate regions according to the learned attention weights ($\alpha_{i,j,k}$), and an identified region by the algorithm is considered correct if it falls into the human-labeled regions. The results by comparing our predicted regions on the ground-truth are shown in Table~\ref{tab:visual-result}. 

It should be noted that selecting top-5 and 10 regions out of 196 candidates itself pose a difficult problem as a ranking task, which is shown by the inferior performance of a randomized selection. By automatic attentive learning over the images, the VECF model gains significant improvements, and by further introducing user reviews as a weak supervision signal, our final Re-VECF model generates much more accurate visual explanations, which verifies that the review information plays an important role in aligning the textual- and visual- features to generate visual explanations.
%by collaboratively learning visual explanations, both VECF and Re-VECF models have very significant improvement from random

\textbf{Qualitative analysis. }
Explainability of recommendations is often assessed qualitatively~\cite{wu2016explaining,ren2017social,wang2011collaborative,he2015trirank}, to provide more intuitive analysis here, we also evaluate our generated visual explanations in a similar manner.
To compare VECF and Re-VECF, we present their generated visual explanations on the same product in the testing dataset, and the parameters follow the default settings as described in section \ref{set}. 
The highlighted regions of a product image are determined by the learned $\alpha_{i,j,k}$ weights in Eq.\eqref{image_att}.
%and we use the category of Clothing/Men as the experimental dataset. 
Examples are presented in Table~\ref{tab:visual-explanation}. From the results we have the following observations.

$\bullet$ Our models can provide meaningful explanations. 
In Case 6, for example, the pocket of the shirt was highlighted by Re-VECF, and in Case 4, VECF labeled the toe of the shoe. These fashion elements are in accordance with the products that the user has purchased.

$\bullet$ In Case 2, although the T-shirts in the historical records are different in many aspects such as color, style, etc, Re-VECF successfully captured their essential similarity -- v-neck, which is highlighted as the visual explanation. This implies the capability of our model to discover users' visual preferences from images and reviews.

$\bullet$ In Case 2 and 6, Re-VECF highlighted different components (collar and pocket) of similar items (shirt) for different users. 
This manifests that our provided visual explanations are personalized, which verifies the effectiveness of our designed user-aware attention mechanism as shown in Eq.\eqref{image_att}.

$\bullet$ By comparing the highlighted regions with the user's reviews, we see that Re-VECF tends to highlight more accurate image regions than VECF. In Case 2 for example, the user praised the collar of the shirt by ``\textit{... this is a really comfortable v-neck ...}'', and Re-VECF successfully labeled the \textit{neck} regions as visual explanation, while VECF highlighted the \textit{sleeve} of the shirt. Other cases also imply the superiority of Re-VECF against VECF in terms of visual explanation. 

%Another example can be seen in Case 3, where the user expressed her favor on the pants' legs, which is also discovered by Re-VisExpCF, but ignored by VisExpCF. 

These observations further verified that the review information leveraged in Re-VECF provides very informative user preferences to better supervise the learning of visual attentions, and thus generates more accurate visual explanations.

\section{Conclusions}
In this paper, we propose \textit{visually explainable recommendation}, aiming to make recommender systems explainable from the visual perspective. To achieve this goal, we proposed two attentive architectures with the supervision of user implicit feedback as well as textual reviews to capture users' visual preferences. Extensive experiments verified that our models were not only able to provide accurate recommendations and review predictions, but also can provide reasonable visual explanations for the recommended items.

This is a first step towards our goal for visually explainable recommendation, and there is much room for further improvements. 
For example, we can integrate probabilistic graphical models with neural modeling to introduce different empirical prior distributions for more accurate visual preference discovery. It will also be interesting to leverage eye-tracking devices to align users' eye attention with the model-learned attention on visual images for visually explainable recommendation.
Beyond e-commerce, we will also investigate visually explainable recommendation in other image-related recommendation scenarios, such as social image recommendation in Instagram or Pinterest, or even multimedia recommendation for stream videos.

%\section*{Acknowledgements}
%We sincerely thank the reviewers for the careful reviews and constructive suggestions.

\bibliographystyle{ACM-Reference-Format}
\bibliography{sample-sigconf}

\end{document}